\documentclass[preprintnumbers,nofootinbib,noshowpacs,eqsecnum,superscriptaddress,prd]{revtex4}

\usepackage{graphicx}
\usepackage{amsmath,amsthm,amssymb}
\usepackage{mathrsfs,bbm}
\usepackage{array,color}

\makeatletter       

\renewcommand{\p@subsection}{}

\makeatother

\newcommand*{\tvec}[1]{\boldsymbol{#1}}              
\newcommand*{\trans}{\mathrm{T}}                     

\DeclareMathOperator{\sgn}{sgn}			

\begin{document}
\title{Scattering amplitudes abandoning virtual particles}

\author{M. Maniatis}
    \email[E-mail: ]{MManiatis@ubiobio.cl}
\affiliation{Departamento de Ciencias B\'a{}sicas, 
UBB, Casilla 447, Chill\'a{}n, Chile.}

\begin{abstract}
We emphasize that scattering amplitudes of a wide class of models to any order in the coupling
are constructible by on-shell tree subamplitudes.
This follows from the Feynman-tree theorem combined with
BCFW on-shell recursion relations.
In contrast to the usual Feynman diagrams,
no virtual particles appear.
\end{abstract}

\maketitle

\section{Introduction}

The success of quantum field theory with respect to the computation of amplitudes
in a perturbative expansion is well-known. Maybe one of the most striking examples
is the computation of the anomalous magnetic moment of the electron \cite{Gabrielse:2006gg}. 

The recent progress in dealing with amplitudes (for recent reviews see for instance 
\cite{Bern:2007dw, Britto:2010xq, Elvang:2013cua}),
in particular, little group scaling and BCFW on-shell-recursion relations suggests that
the actual computation of a physical amplitude follows from a rather small number
of subdiagrams or masteramplitudes. One example is the Parke-Taylor formula \cite{Parke:1986gb}
for the lowest order $n$-gluon amplitude which follows via recursion relations from
the simple three-gluon amplitude. 

Here we want to emphasize that amplitudes of a given loop coupling order in general may be construced
by  on-shell tree subamplitudes. This observation is based on a
combination of the Feynman-tree theorem 
\cite{Feynman:1963ax, Feynman:FTT} with BCFW on-shell-recursion relations of
tree diagrams \cite{Britto:2004ap, Britto:2005fq}. In particular, there do not appear
virtual particles in this way. This in turn makes the use of ghosts
obsolete. In this picture all subamplitudes are on-shell, but we have to deal with {\em hidden} particles,
that is, external background on-shell particles which are unobserved. The method works in any
 theory in which the boundary term of the BCFW recursion relations vanishes. This was shown to
 hold in gauge theories as well as in general relativity \cite{ArkaniHamed:2008yf}.
Any amplitude is constructed by merging on-shell subamplitudes together, where in general we encounter
a background of external {\em hidden} particles. 

Let us mention the recent interest in the Feynman-tree theorem; see
for instance  \cite{Brandhuber:2005kd, Catani:2008xa, CaronHuot:2010zt, Boels:2010nw, Bierenbaum:2010cy}.
One of the strategies is to reduce the in general large number of tree amplitudes
avoiding multiple cuts.
Here we will apply
the original version of the Feynman-tree theorem.

\section{The Feynman-tree theorem and on-shell recursions}

\begin{figure}[Ht!]
\includegraphics[width=0.3\textwidth]{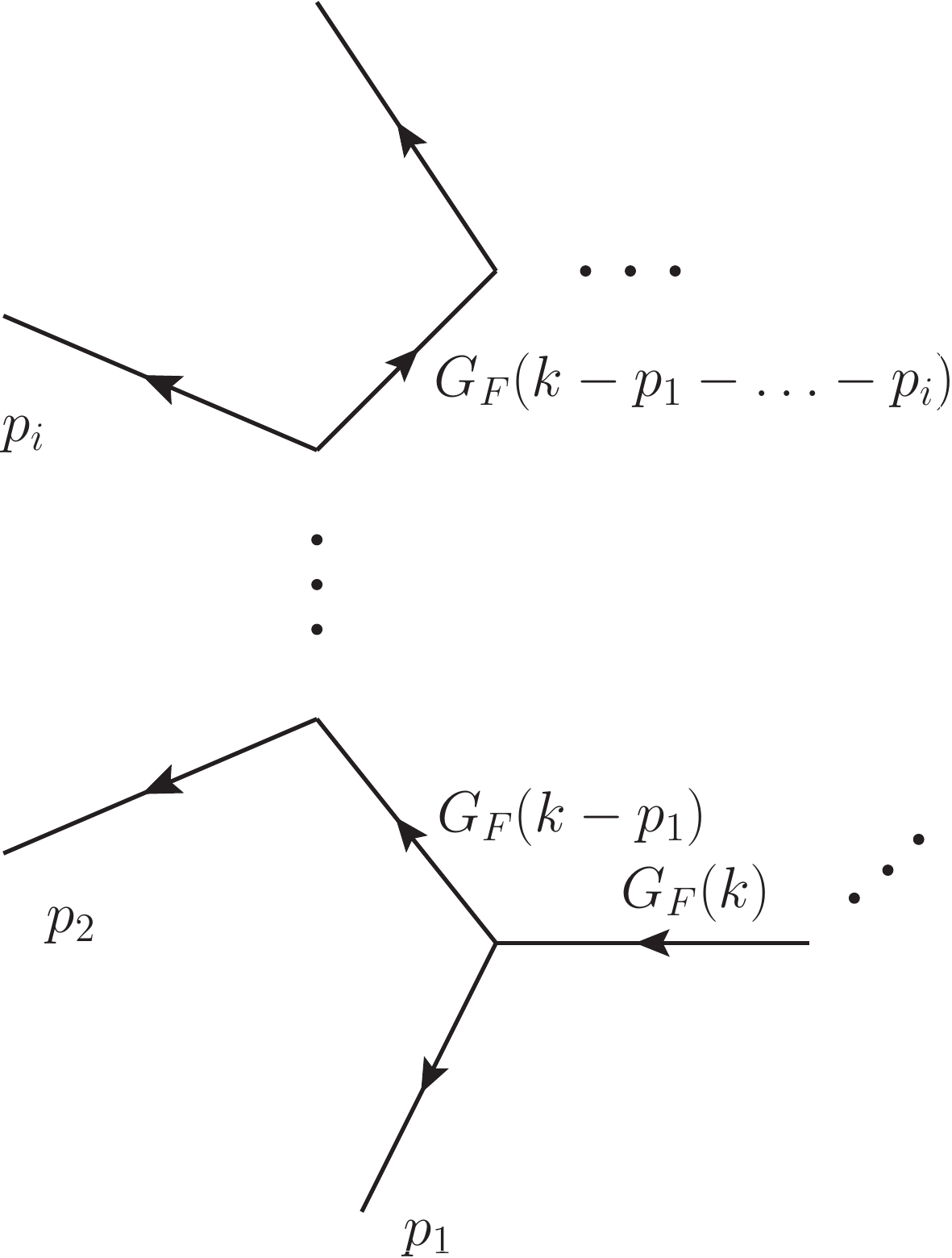}
\caption{\label{loopFTT}A generic loop in a diagram with loop momentum $k$ and 
the usual Feynman propagators denoted
by $G_F$. All leg momenta are chosen by convention outgoing, 
as indicated by the little arrows. The outgoing legs may correspond to one or more particles and
the indicated momentum is the sum of the outgoing momenta.
}
\end{figure}

Let us briefly review the Feynman tree theorem \cite{Feynman:1963ax, Feynman:FTT},  
which reduces $l$ loop amplitudes to at most $l-1$ loop amplitudes systematically, that 
is, recursively to tree amplitudes.
The basic idea is to introduce besides the usual propagators $G_F(p)$ also 
{\em advanced} propagators $G_A(p)$
\begin{equation} \label{GA}
G_F(p) = \frac{i}{p^2 - m^2 + i \epsilon}, \qquad
G_A(p) = \frac{i}{p^2 - m^2 - i \epsilon \sgn(p_0)}.
\end{equation}
From the identity
\begin{equation}
\frac{1}{x\pm i \epsilon} = P.V. \bigg(\frac{1}{x}\bigg) \mp i \pi \delta(x),
\end{equation}
where $P.V.$ denotes the principal value prescription, 
we get a simple context between the usual propagators and the advanced ones,
\begin{equation} \label{GAGF}
G_A(p) = G_F(p) - 2 \pi \theta(p_0) \delta(p^2 - m^2).
\end{equation}
Let us consider a generic loop diagram as shown in Fig. \ref{loopFTT}. Here $k$ 
denotes the loop momentum of the considered loop and all momenta $p_i$ with $i=1,\ldots,n$ are chosen by convention
to be outgoing. We note that the outgoing momenta in this diagram may also correspond to two particles in each vertex. 
In these cases the indicated momentum gives the sum of the outgoing momenta.
We note further that the legs do not need to be external. In this sense, the loop diagram in Fig. \ref{loopFTT} is generic.

In the considered loop of a diagram we replace all Feynman propagators $G_F(p)$ by
the advanced propagators $G_A(p)$.
In the integration over the time-like component
of the loop-four momentum, $k_0$, all poles of the advanced propagators lie now above the real axis.
Closing the contour in the lower half plane, the loop integral is zero. Using \eqref{GAGF} this gives
\begin{equation} \label{FTT}
\begin{split}
0 =& \int \frac{d^4 k}{(2\pi)^4}  N(k)\; \prod_i G_A^{(i)}(k-p_1-\ldots-p_i)\\ 
=& 
\int \frac{d^4 k}{(2\pi)^4}  N(k) \; \prod_i \bigg\{ G_F^{(i)}(k-p_1-\ldots-p_i) -
2 \pi \theta(k_0-p_{10}-\ldots-p_{i0}) \delta((k-p_1-\ldots-p_i)^2 -m^2) \bigg\}.
\end{split}
\end{equation}
Here $N(k)$ denotes the numerator of the amplitude, which in general also depends on the 
loop-momentum $k$.
The product in the last line of \eqref{FTT} is the desired recursion relation of the 
loop amplitude: one term of this expansion is the original loop amplitude 
with all Feynman propagators
and all remaining terms with one or more propagators replaced by the  
corresponding delta-function terms.
The delta-function term together with the loop momentum integration
corresponds to a phase-space integration with the cut propagators on-shell. In general
a loop with $n$ propagators gives $2^n-1$ cut diagrams.
The Feynman-tree recursion gives a sum of new amplitudes with the loop order decreased 
about at least one unit in each recursion step. 
Repeated application of this recursion relation represents all loop amplitudes in terms of
tree amplitudes.

Now we emphasize that by general BCFW tree-recursion relations \cite{Britto:2004ap,Britto:2005fq} we can 
express the tree amplitudes resulting from the Feynman-tree theorem, in terms of on-shell amplitudes. 
The basic idea of the BCFW recursion relations 
is analytic continuation of the external momenta. In this way 
tree amplitudes factorize into on-shell subamplitudes.
In an arbitrary tree amplitude let us denote the $n$ external momenta
by $p^\mu_i$ with $i=1,\ldots,n$.
These external momenta are shifted,
\begin{equation}
\hat{p}_i^\mu = p_i^\mu + z \cdot r_i^\mu
\end{equation}
with one common $z \in \mathbb{C}$ and appropriately chosen vectors $r_i$. 

\begin{figure}[Ht!]
\includegraphics[width=0.6\textwidth]{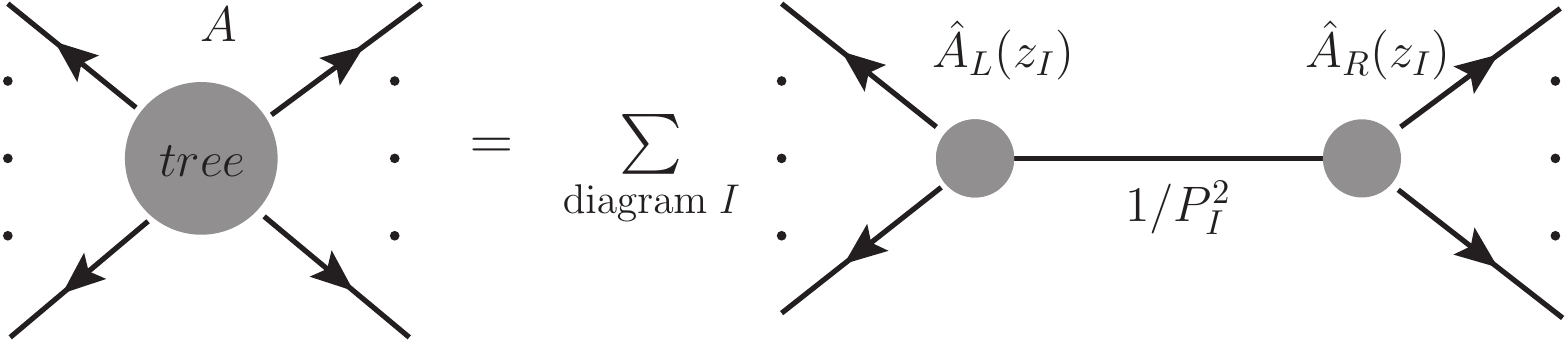}
\caption{\label{onshell}BCFW on-shell recursion relation in case of a 
vanishing boundary term $B$. The tree amplitude $A$ is decomposed into a sum
of on-shell subdiagrams with left and right parts $\hat{A}_L$ and $\hat{A}_R$ and
propagator factor $1/P_I^2$.
}
\end{figure}

The statement is that any tree amplitude $A$ can by analytic continuation be decomposed in
terms of on-shell subamplitudes connected by propagators and a boundary term $B$,
\begin{equation}
A = - \sum_{z_I} \text{Res}_{z=z_I} \frac{\hat{A}(z)}{z} + B =
\sum_{\text{diagram }I} \hat{A}_L(z_I) \cdot \frac{1}{P_I^2} \cdot \hat{A}_R(z_I) + B.
\end{equation}
Here, $\hat{A}(z)$ denotes the shifted amplitude with the positions of poles in the complex
plane at $z=z_I$. On the right-hand side we
have the on-shell subamplitudes $\hat{A}_L(z_I)$ and $\hat{A}_R(z_I)$, with a propagator factor $1/P_I^2$.
In general, there appears also a boundary term $B$, which is the residue of the pole of
$\hat{A}(z)$ at $z=\infty$. In case of a vanishing term $B$ we have on the right-hand side
the desired factorization into on-shell subamplitudes, as shown 
in Fig. \ref{onshell}.

It has been shown that for an appropriate shift of the external momenta in gauge theories
as well as general relativity
the boundary term $B$ vanishes \cite{ArkaniHamed:2008yf}. 
We have, therefore in these cases a recursion
of tree amplitudes to on-shell amplitudes. An example of a theory, where we do 
not have a vanishing boundary contribution $B$ is $\phi^4$ theory, as discussed in \cite{Feng:2009ei}.

Combining the two recursion methods, that is, on the one hand the Feynman-tree theorem
to transform loop amplitudes to tree amplitudes and subsequently BCFW recursions to transform
the tree amplitudes
to on-shell amplitudes we can express amplitudes in terms of on-shell subamplitudes. 
Obviously, some of the external particles are {\em hidden}, originating from the Feynman-tree theorem.

We emphasize that we equivalently can start the construction of
amplitudes by on-shell subamplitudes. This holds when the boundary terms vanish. 
In this picture amplitudes are constructed by
joining on-shell subamplitudes to a given order in the coupling. This gives
a new interpretation, which, as we have seen, is equivalent to the conventional Feynman diagram approach.
Respecting the Feynman-tree theorem, we have in general to consider
amplitudes with {\em hidden} external particles, that is, particles which are unobserved.
In this picture virtual particles do not appear. In particular, no ghosts have to be introduced.
Moreover, it is clear that every subdiagram
is on-shell, that is, in particular, gauge invariant. This is to compare with
the usual Feynman diagram construction of amplitudes where in general gauge invariance
is violated in each diagram but is only restored in the sum of diagrams to a given order.
An explicit example of a two-point amplitude, to second order in the coupling, is studied in the appendix. We have in this example
to consider three on-shell amplitudes. Conventionally, these
correspond to a one-loop Feynman diagram.

\section{Conclusions}

We have emphasized that amplitudes can be constructed by on-shell subamplitudes.
This works in theories where the corresponding BCFW recursion relations have a vanishing boundary term.
It has been shown that this holds in gauge theories as well as general relativity.
All the poles and branch cuts
follow automatically in this picture. Consequentially, there are no
virtual particles and no ghosts in this picture. This construction of gauge invariant 
on-shell amplitudes follows directly
from the Feynman-tree theorem combined with BCFW on-shell recursion relations.
To a given order in the couplings, amplitudes with a background of external, but {\em hidden} particles
have to be introduced. We have to integrate over the corresponding phase space of the {\em hidden} particles.
Let us mention that in cases where we can construct the lowest order on-shell amplitudes
by little group scaling, that is, eventually Lorentz invariance, 
all amplitudes follow recursively by this construction.

\section*{Acknowledgement}
We would like to thank 
Carlos Reyes and in particular Otto Nachtmann
for many valuable comments and suggestions.
This work is supported partly by the 
Chilean research project FONDECYT, with project number
1140568 as well as by the group of {\em F\'{i}sica
de Altas Energias} of the Universidad del B\'{i}o--B\'{i}o.

\appendix
\section{A two-point function example}

Let us consider as a simple example a scalar theory with Lagrangian

\begin{equation} \label{phi3}
{\cal{L}} = \frac{1}{2} (\partial_\mu \phi) (\partial^\mu \phi) - \frac{1}{2} m^2 \phi^2 - \frac{g}{3!} \phi^3 .
\end{equation}

We want to compute the 2-point amplitude at order $g^2$.
We start with the conventional Feynman diagram computation, giving\\
\centerline{
\begin{tabular}{m{36pt}m{24pt}m{100pt}}
$-i A(p^2)$ 
& $=$ &
\includegraphics[width=120pt]{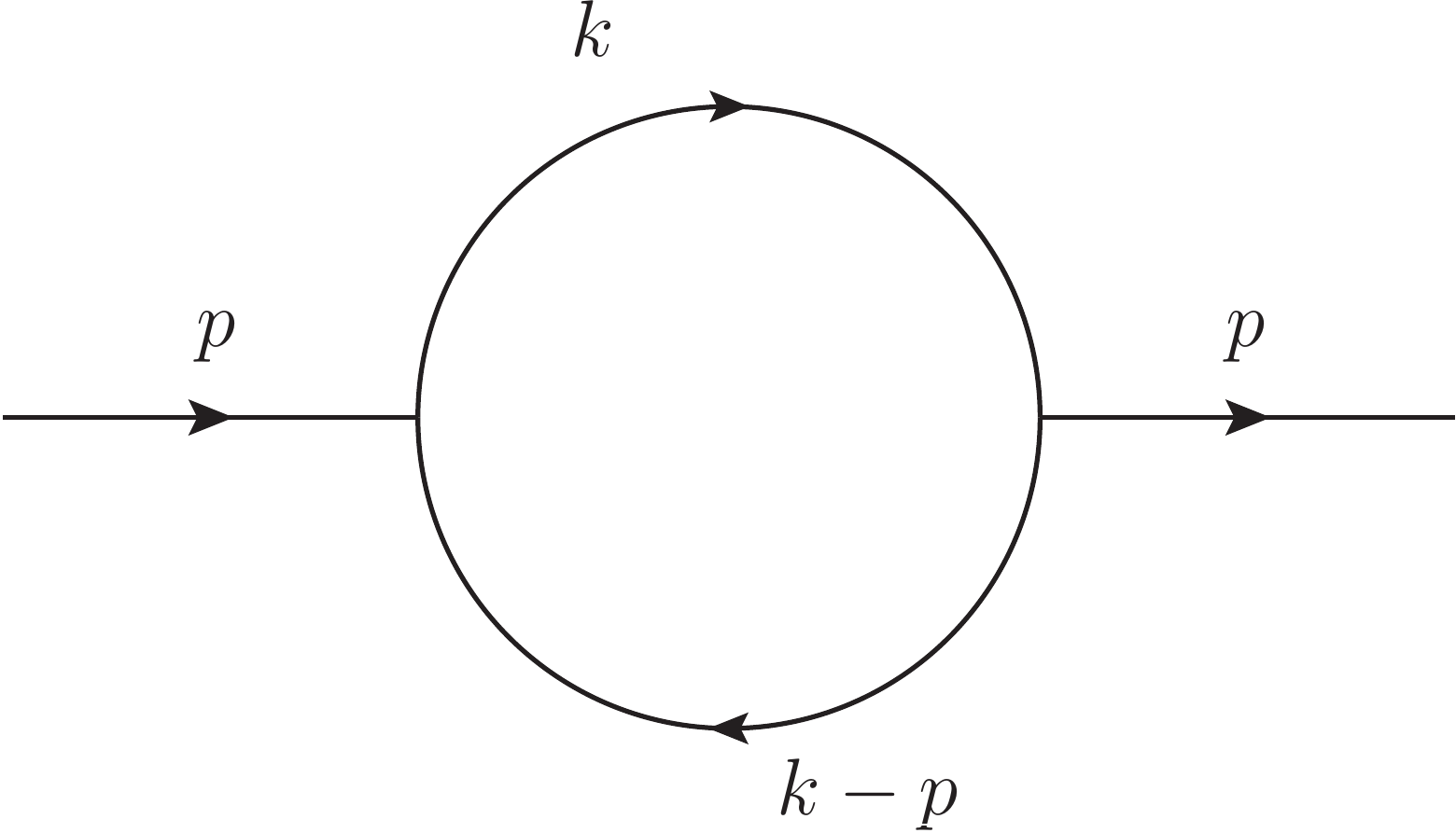}
\end{tabular}
}
We have
\begin{equation} \label{selfFeyn}
-i A(p^2) = \frac{(-ig)^2}{2} \int \frac{d^4 k}{(2\pi)^4} 
\frac{i}{k^2-m^2+i \epsilon} \cdot
\frac{i}{(k-p)^2-m^2+i \epsilon}. 
\end{equation}
After Feynman parametrization we regularize the UV divergence for simplicity 
with a heavy mass $M \gg m$, that is, Pauli-Villars regularization,
\begin{equation}
-i A^{\text reg}(p^2) = \frac{(-ig)^2}{2}  \int_0^1 dx \int \frac{d^4 k}{(2\pi)^4} 
\big\{ k^2 +p^2x^2 - p^2x -m^2\big\}^{-2} - \big\{ k^2 +p^2x^2 - p^2x -M^2\big\}^{-2} .
\end{equation}
As usual, going to Euclidean space and utilizing spherical coordinates we end up with
\begin{equation}
\begin{split}
-i A^{\text reg}(p^2) =& i \frac{g^2}{32 \pi^2} \int_0^1 dx \ln  \big\{ \frac{m^2-p^2x-p^2x^2}{M^2} \big\}\\
=& i \frac{g^2}{32 \pi^2} \bigg\{ 2 \sqrt{\frac{4m^2}{p^2}-1} \arctan \big( \sqrt{ \frac{p^2}{4m^2-p^2} } \big)
- 2 - \log \big( \frac{M^2}{m^2} \big) \bigg\}.
\end{split}
\end{equation}

Now we want to show that this is equivalent to on-shell amplitudes to the same order $g^2$.
In this we have to consider also amplitudes with {\em hidden} external particles,
that is,

\begin{tabular}{m{36pt}m{6pt}m{144pt}m{6pt}m{144pt}m{6pt}m{144pt}}
$-i A(p^2) = $ 
& &
\includegraphics[width=144pt]{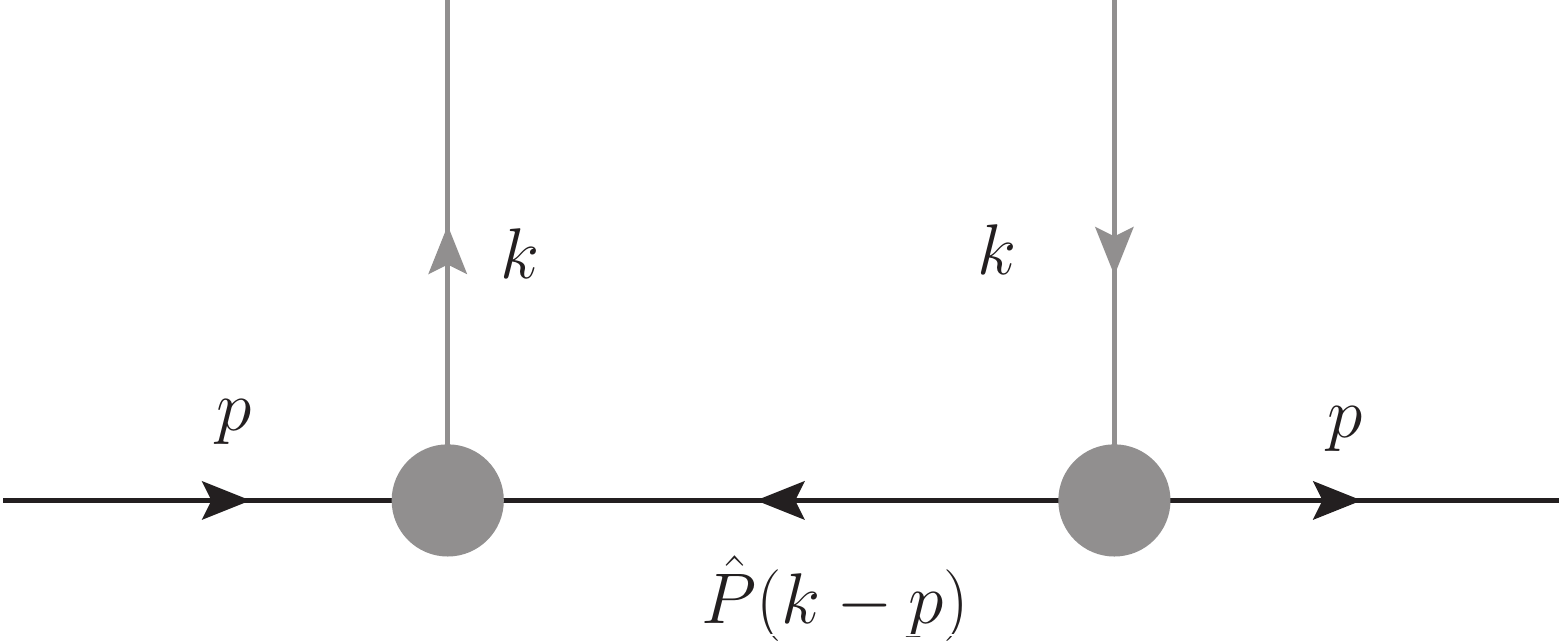}
& + &
\includegraphics[width=144pt]{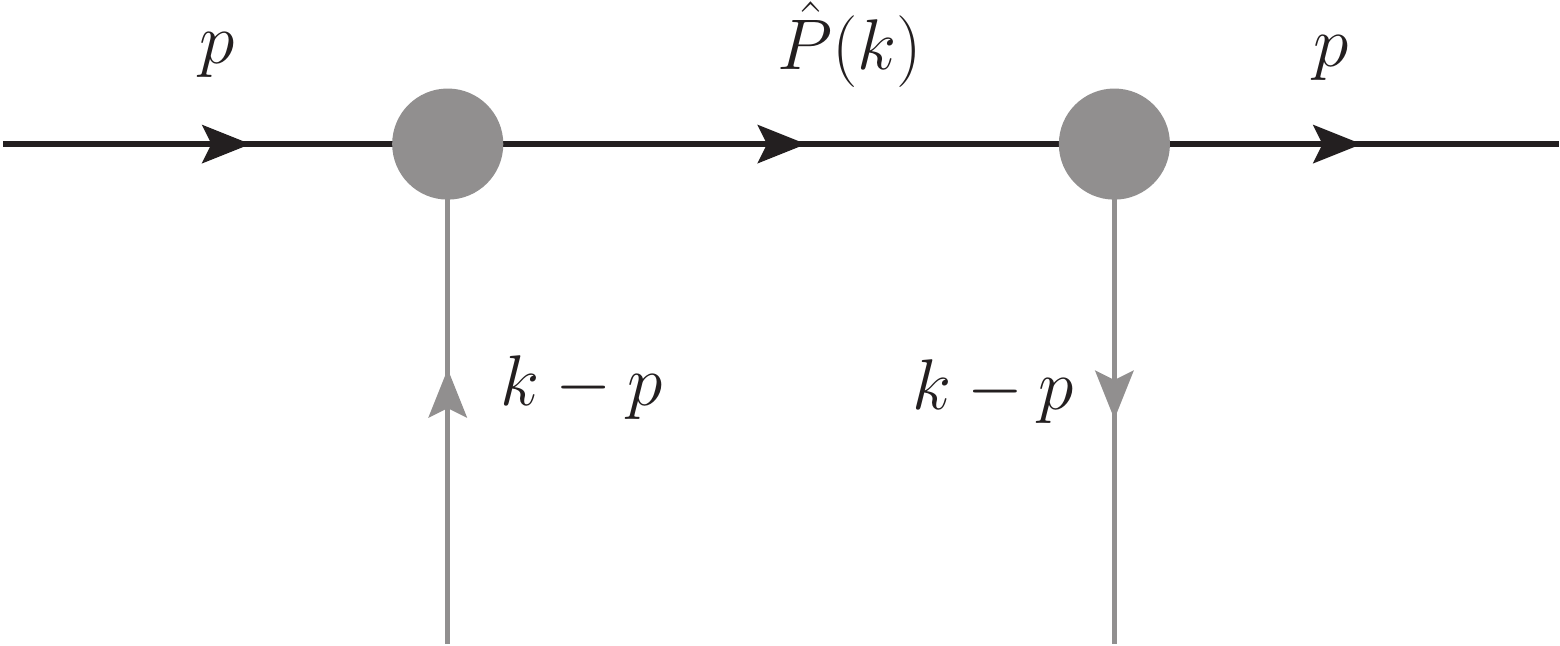}
& + &
\includegraphics[width=144pt]{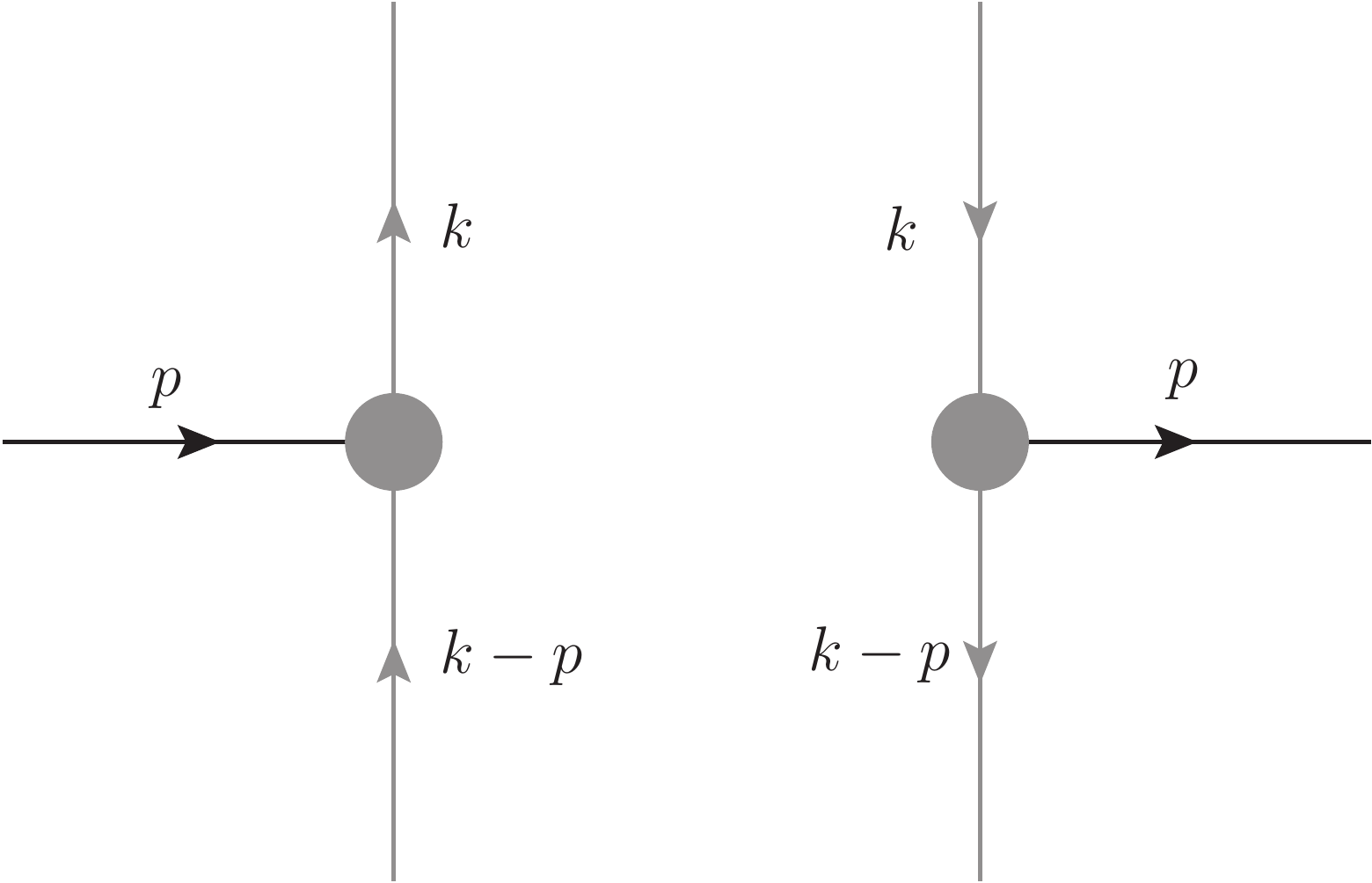}
\end{tabular}\\
All vertical lines denote in these amplitudes the {\em hidden} external particles, that is, we
have to integrate over the corresponding phase space.\\

To show the equivalence we start with 
\eqref{selfFeyn} and
 apply the Feynman-tree theorem, replacing the Feynman propagators $G_F$ by advanced propagators $G_A$ \eqref{GAGF}, 
 yielding ($p = (p_0, \tvec{p})^\trans$ and
 $\omega_{\tvec{p}} = +\sqrt{\tvec{p}^2+m^2}$).  
\begin{equation} \label{selfFTT}
\begin{split}
-i A(p^2) = \frac{(-ig)^2}{2} \int \frac{d^4 k}{(2\pi)^4} 
\bigg\{&
G_A(k-p) \frac{\pi}{\omega_{\tvec{k}}} \delta (k_0 - \omega_{\tvec{k}} )\\
& +
G_A(k) \frac{\pi}{\omega_{\tvec{k}-\tvec{p}}} \delta (k_0 - p_0 - \omega_{\tvec{k}-\tvec{p}} )\\
& +
\frac{\pi}{\omega_{\tvec{k}}} \delta (k_0 - \omega_{\tvec{k}}) \cdot
\frac{\pi}{\omega_{\tvec{k}-\tvec{p}}} \delta (k_0 - p_0 - \omega_{\tvec{k}-\tvec{p}} )
\bigg\}.
\end{split}
\end{equation}
These are all possible cuts we can apply to the loop, diagramatically

\begin{tabular}{m{36pt}m{6pt}m{144pt}m{6pt}m{144pt}m{6pt}m{144pt}}
$-i A(p^2) = $ 
& &
\includegraphics[width=144pt]{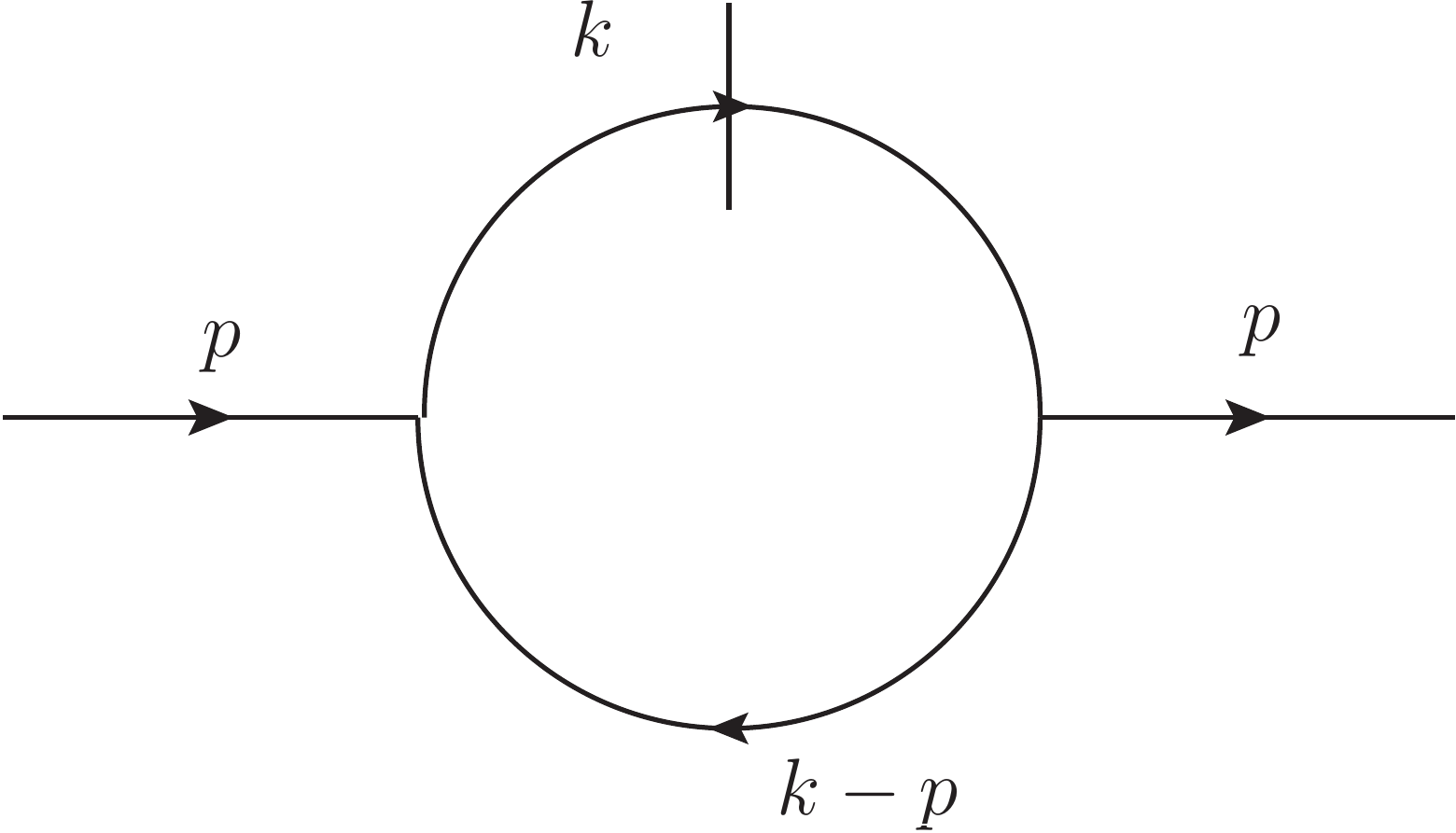}
& + &
\includegraphics[width=144pt]{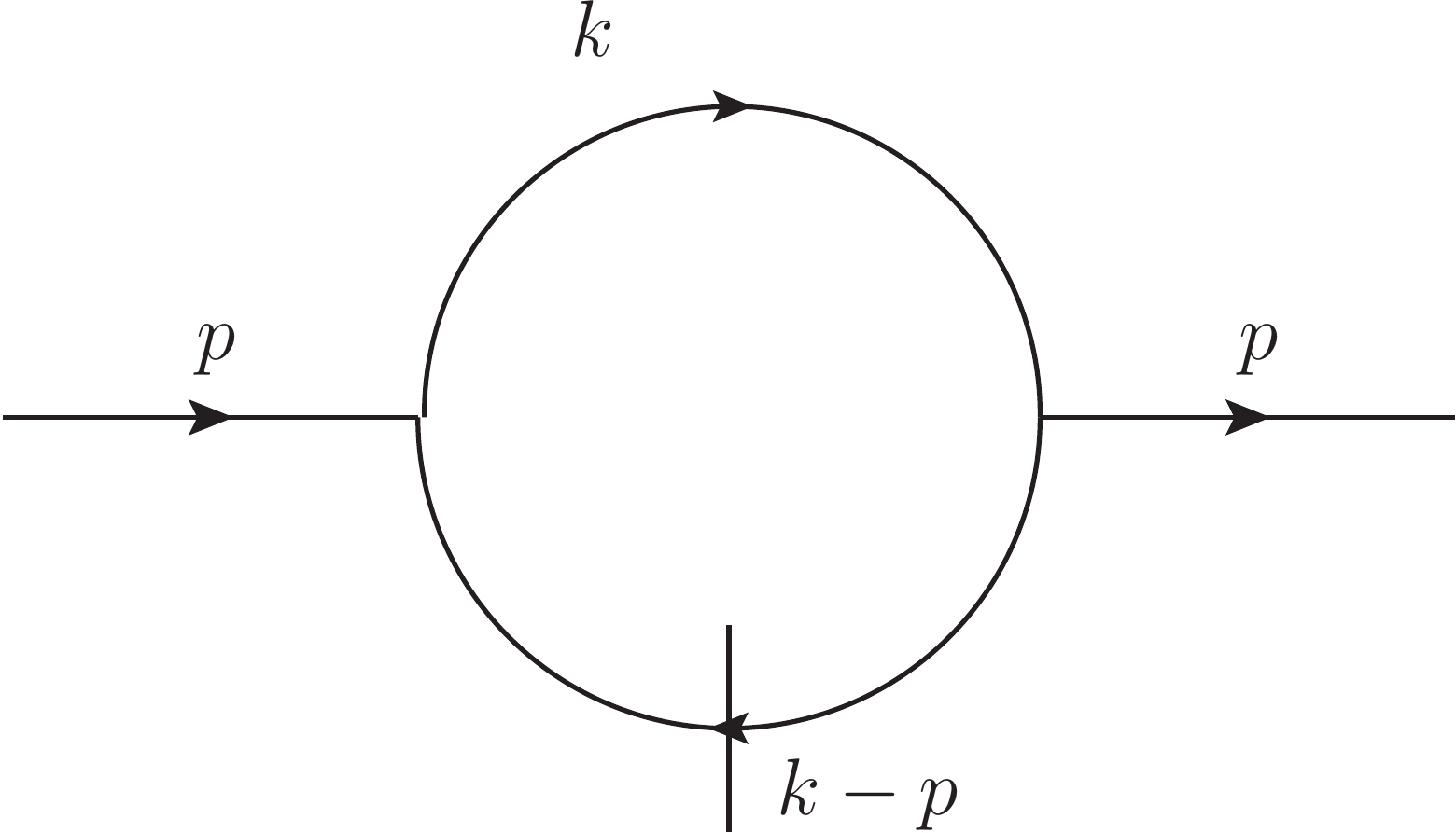}
& + &
\includegraphics[width=144pt]{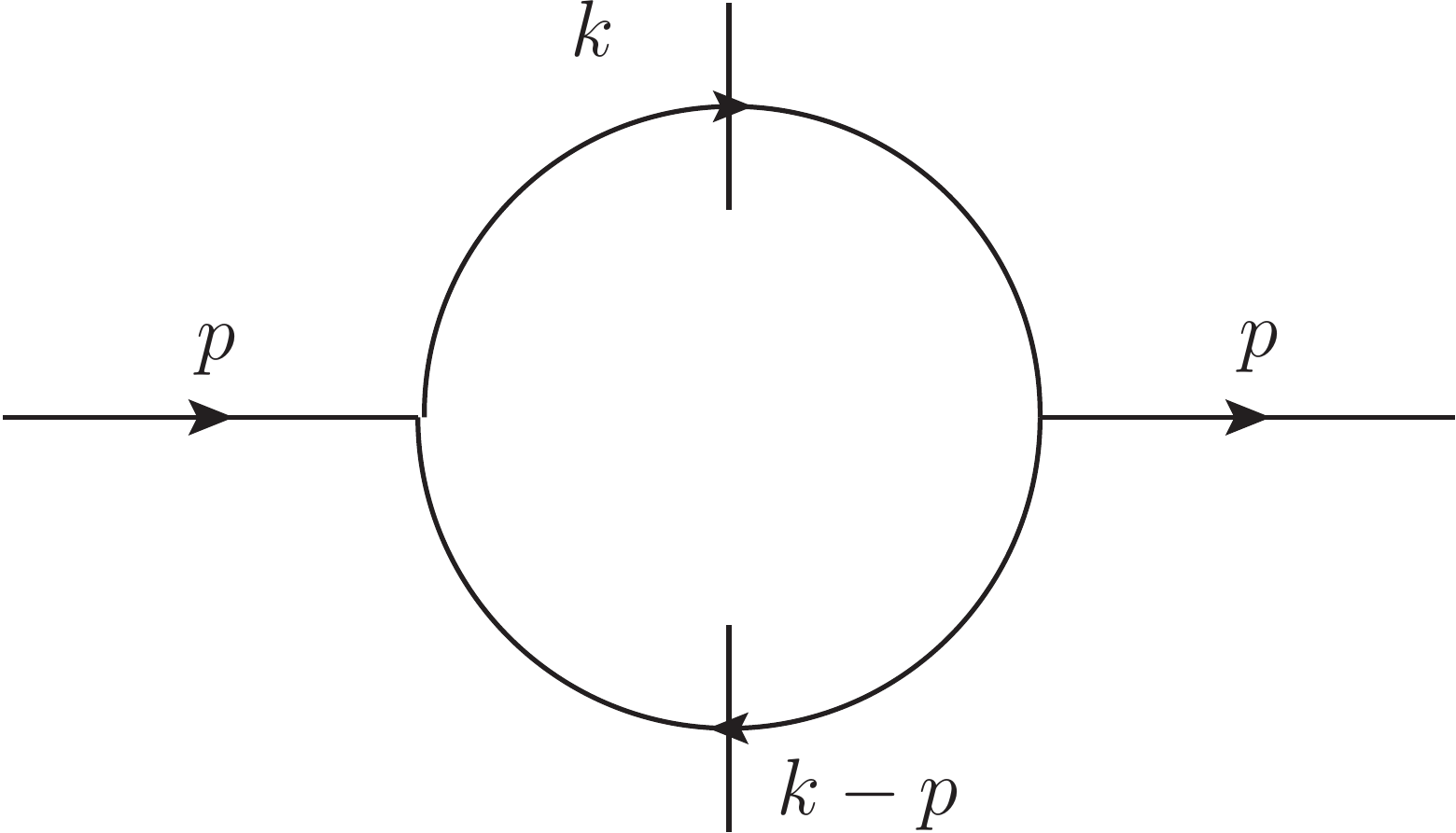}
\end{tabular}\\

Let us consider the first term in \eqref{selfFTT}. corresponding to the first diagram.
We have
\begin{equation} \label{self1}
\begin{split}
-i A(p^2)_{\text{first}} = &
\frac{(-ig)^2}{2} \int \frac{d^4 k}{(2\pi)^4} 
G_A(k-p) \frac{\pi}{\omega_{\tvec{k}}} \delta (k_0 - \omega_{\tvec{k}} )
\\
= &
\frac{(-ig)^2}{2} \int \frac{d^3 k}{(2\pi)^4} 
\frac{\pi}{\omega_{\tvec{k}}} 
\frac{i}{ (\omega_{\tvec{k}} - p_0 - i \epsilon)^2 - (\tvec{k}-\tvec{p})^2 - m^2 }
\\
= &
\frac{(-ig)^2}{2} \int \frac{d^3 k}{(2\pi)^3 2 \omega_{\tvec{k}}} 
\frac{i}{ (k-p)^2 - m^2 } .
\end{split}
\end{equation}
Since in this example we deal with scalars only, the last line in \eqref{self1}
gives already the first on-shell amplitude above.

In the rest frame of $p$ we get
\begin{equation} \label{selfc1}
-i A(p^2)_{\text{first}} = 
\frac{(-ig)^2}{2} \int \frac{d^3 k}{(2\pi)^3 2 \omega_{\tvec{k}}} 
\frac{i}{ p^2- 2 p \omega_{\tvec{k}} } .
\end{equation}
Similar we get for the second cut diagram
\begin{equation} \label{selfFTT2}
-i A(p^2)_{\text{sec.}} = 
\frac{(-ig)^2}{2} \int \frac{d^3 k}{(2\pi)^3 2 \omega_{\tvec{k}-\tvec{p}} }
\frac{i}{ k^2 - m^2 } =
\frac{(-ig)^2}{2} \int \frac{d^3 k}{(2\pi)^3 2 \omega_{\tvec{k}}} 
\frac{i}{ p^2 + 2 p \omega_{\tvec{k}} } 
\end{equation}
which is the second on-shell amplitude above. 
In the $p$ rest frame this corresponds to the replacement $p \to -p$ in \eqref{selfc1}.

The third term in \eqref{selfFTT} does not contribute. This follows from the fact that the two delta functions
are never simultaneously satisfied for $p_0 > \tvec{p}$, since
$p_0 + \sqrt{(\tvec{k}-\tvec{p})^2+m^2} > \sqrt{\tvec{k}^2+m^2}$.
That is we have 
\begin{equation}
-i A(p^2) = 
-i A(p^2)_{\text{first}}
-i A(p^2)_{\text{sec.}}
\end{equation}

We regularize the UV divergence, subtracting a heavy mass $M \gg m$ term and have
\begin{equation} \label{selfc2}
-i A^{\text{reg}}(p^2) = 
\frac{(-ig)^2}{2} \int \frac{d^3 k}{(2\pi)^3}
 2 \omega_{\tvec{k}} 
\bigg\{
\frac{i}{ 2 \sqrt{\tvec{k}^2+m^2}\cdot (p^2- 2 p \sqrt{\tvec{k}^2+m^2}) } 
-
\frac{i}{ 2 \sqrt{\tvec{k}^2+m^2}\cdot (p^2- 2 p \sqrt{\tvec{k}^2+M^2}) } 
\bigg\} .
\end{equation}

In spherical coordinates the integration gives
\begin{equation}
-i A^{\text reg}(p^2) = 
\frac{(-ig)^2}{2} 4 \pi \frac{1}{8p}
\bigg\{
2 \sqrt{4m^2-p^2} \arctan \big( \frac{p}{\sqrt{4m^2-p^2}} \big) 
-
2 \sqrt{4M^2-p^2} \arctan \big( \frac{p}{\sqrt{4M^2-p^2}} \big)
+ p \log{\frac{m^2}{M^2}}
\bigg\}
\end{equation}
what in the limit of $M^2 \gg m^2$ is the same as what we found with the usual Feynman diagram calculation.

\end{document}